\documentclass[11pt]{article}
\usepackage{epsfig,latexsym}

\newcommand{\proof}{{\noindent \bf Proof. }}

\newtheorem{thm}{Theorem}[section]

\newtheorem{cor}{Corollary}[section]
\newtheorem{defi}{Definition}[section]

\newtheorem{lem}{Lemma}[section]

\def\sF{{\cal F}}
\def\sG{{\cal G}}
\def\sD{{\cal D}}

   \def\polylog{{\rm poly}\log }

\def\sS{{\cal S}}
\def\sR{{\cal R}}

    \newcommand{\sP}{{\cal P}}
\newcommand{\sA}{{\cal A}}

\def\phase{\mbox{\sc phase}}
\def\stage{\mbox{\sc stage}}

\newcommand{\Cnk}[2]{\left( \begin{array}{c} #1 \\ #2 \end{array}%
\right)}

\newcommand{\prot}[1]{\mbox{{\sc broad}$^{(#1)}$}}
\newcommand{\prota}{\mbox{\sc broad-a}}
\newcommand{\protb}{\mbox{\sc broad-b}}

\newcommand{\allprota}{\mbox{\sc multi-bb-broad-a}}

\newcommand{\eproof}{\hfill$\Box$  \medskip}

\begin{document}

\title{{
\bf
   Distributed  Broadcast   in   Wireless  Networks
   with Unknown Topology\thanks{Research
   partially supported   by the European RTN Project
{\em ARACNE} and by the Italian MURST Project
``Resource Allocation in Communication
                      Networks". The results of this paper have been
                      presented at {\em ACM-SIAM  SODA'01} \cite{CMS01} and at {\em
                      ACM PODC'01} \cite{CMS01b}.}
  }}

\author{  Andrea E. F.  Clementi\thanks{
Dipartimento di Matematica,
Universit\`a di Roma "Tor Vergata",
    Email: {\tt
clementi@mat.uniroma2.it. }}
   \and
   Angelo Monti\thanks{Contact Author:
   Dipartimento di Scienze dell'Informazione,
   Universit\`a ``La Sapienza'' di Roma, Email:   {\tt
monti@dsi.uniroma1.it}.}
   \and
   Riccardo Silvestri\thanks{
   Contact Author:
   Dipartimento di Scienze dell'Informazione,
   Universit\`a ``La Sapienza'' di Roma,  
   Email: {\tt silver@dsi.uniroma1.it}.}
   }
\date{\today}

\bibliographystyle{alpha}


\maketitle
\thispagestyle{empty}

\begin{abstract}
   A multi-hop synchronous wirelss network is said to be {\em unknown}
   if the nodes have no knowledge of the topology.  
   A basic task in wireless network
   is that of {\em broadcasting} a message (created by a fixed source node)
    to all nodes of the  network. Typical operations in real-life 
    wireless networks
is the {\em multi-broadcast} that
   consists in performing a set of $r$ independent  broadcasts.
   The study of broadcast operations on unknown wireless network 
   is started by the seminal paper of  Bar-Yehuda
{\em et al}~\cite{BGI92} and has been the subject of several  recent  
works.

   \noindent
   In this paper, we study the completion and the termination time
   of distributed protocols  for  both the 
   (single)  broadcast   and   the
   multi-broadcast operations on unknown
   networks as functions of the number of   nodes $n$, the maximum eccentricity $D$,
   the  maximum in-degree $\Delta$, and  the congestion $c$ of the    networks.
    We establish new
   connections between these operations and some  
   combinatorial concepts, such as {\em selective families}, {\em 
   strongly-selective families} (also known as {\em 
   superimposed codes}), and {\em pairwise $r$-different families}.
   Such connections, combined with a set of new lower  and upper
   bounds on the size of the above families,  allow us to derive 
     new lower bounds and new   distributed protocols for  the  
     broadcast and multi-broadcast operations.
     
     \noindent
     In particular, our 
     upper bounds   are almost tight and      improve exponentially  over the
       previous bounds  when 
     $D$ and $\Delta$ are polylogarithmic in $n$. Network topologies having  ``small''
     eccentricity and ``small'' degree (such as bounded-degree {\em 
     expanders}) are often used in practice to achieve efficient
     communication.

\end{abstract}
\newpage

   \section{Introduction}\label{sec::intro}

\subsection{Wireless networks}\label{sec::int-rn}

{\em Static ad-hoc wireless networks} (in short, radio networks)
  have been the subject of several works in recent years due to
  their potential applications in scenarios such as
  battlefields, emergency disaster relief,
  and  in   any situation in which it is very difficult (or impossible) to
  provide the necessary infrastructure \cite{Rap96,WNE00}.
  As in other network models,   a challenging task is to enable  fast
   communication.

A {\em  radio network}  is a set of
radio stations  that are able to
communicate by transmitting  and receiving  radio signals. A
transmission range is assigned  to  each station $s$ and any other
station    within this range can directly (i.e. by one {\em
hop}) receive  messages from $s$. Communication between two
stations that are not within their respective ranges can be
achieved by   {\em multi-hop} transmissions.    A useful (and sometimes
unavoidable) paradigm of radio communication is the structuring of
communication into synchronous {\em time-slots}. This paradigm is
commonly adopted in the practical design of protocols and hence the use
of the paradigm in theoretical analysis is well motivated
\cite{BGI92,G85,R72}.

   A radio network can be modeled as a directed graph   where an
edge $(u,v)$  exists if and only if     $u$ can communicate with  $v$ in
one hop. The nodes of  a radio network are processing units,
each of which is  able to perform local   computations. It is also assumed
that every node can  perform {\em any\/}    local computation
required for deciding the next send/receive operation   during
the current time-slot.
   In every time-slot, each node can be {\em active} or {\em non
active}. When   active, it
   can decide to be either   {\em transmitter} or
{\em receiver}: in the former case the node transmits
   a message along all of
its outgoing edges while, in the latter case, it
tries to recover messages from all its incoming edges. When it is not
active, it does not perform any kind of operation. The fundamental
feature here is that a node $v$ can recover a message from one of its
incoming edges if and only if this edge is the only one bringing in a
message. If two or more neighbors of a node    are transmitting
   at the same time-slot then a {\em collision}
   occurs. Nodes do not distinguish between the background noise and the
interference noise (i.e., we are in the case of absence of {\em collision
detection} \cite{PL95,BGI92}).
A radio network is said to be
{\em unknown} when    every   node
     knows   nothing about the network but  its  own
     label~\cite{BGI92,CGGPR00,CGOR00,CGR00}.  Informally speaking,
unknown radio networks, with absence of collision detection,
model communication networks with multi-access channels
in which the assumptions on the processors' knowledge   and on the channel
    are minimal.
   In fact, in several applications, the network topology is unstable or
   dynamic  and it is difficult to distinguish the presence of a collision
   from the background noise of the channel.
   Another important motivation
in studying       unknown networks comes from its strong connection
with the {\em fault-tolerance} issue \cite{PR97, KKP98, KM98,CMS01c}.
Unknown networks  can also  be    seen
        as ``known'' networks (i.e. networks in which nodes have the
   knowledge of the entire initial topology)
   with   {\em unknown permanent  faults}. A node (or an edge) is said
   to suffer a permanent fault if it is never active during the entire
   execution of the  protocol \cite{KKP98, CMS01c}. It is easy to show
   that  a broadcast protocol for
   unknown fault-free networks is also an (unknown permanent-)fault 
tolerant protocol
   for general (known) networks   and
   viceversa. So,   the results obtained in the unknown model
   immediately apply   to the permanent-fault tolerance issue.

\medskip

One of the fundamental tasks in network communication is
    the  {\em broadcast} operation. It
   consists in transmitting a   message
from one   source node to all the nodes.

     According to the network model
described above, the communication protocol    operates in
{\em    time-slots}:
    at every time-slot, each
active node decides to either transmit or receive, or turn into
   the non active
state.
Two kinds of broadcast protocols have been considered in the literature
\cite{BGI92,BD97, CGGPR00,CGOR00}:
{\em  spontaneous protocols}, in which   the starting time-slot is known
to all the nodes and, thus,  every node can transmit even if it has not
received any message in     previous time-slots; {\em non spontaneous
   protocols}
in which a node (which is not the source)
may act as a transmitter in a time-slot only if it has
received a message in some previous time-slots (while the source 
starts at time-slot 0).
A deterministic (randomized) broadcast protocol  is said to have  {\em 
completed broadcasting}
  when all
nodes, reachable from the source, have received (with high
probability\footnote{A formal definition of completion time for
randomized protocols will be given later.}) the source message.
Notice that when this happens, the
nodes not necessarily  stop to run the protocol since they might not know
that the operation is completed.
We also say that a
broadcast protocol {\em terminates} in time $t$ if, after the time-slot
$t$,
all the nodes are in the non active state
    (i.e. when all nodes stop to run the protocol).

The completion and termination time of
  {\em Deterministic (Randomized)
   Broadcast protocols}, in short  DB (RB) protocols,  will be  analysed
   as   functions of the following parameters of the
    network: the number $n$ of nodes, the maximum in-degree $\Delta$,
   and the  maximum eccentricity $D$ over all possible source nodes. Given a
   source node $s$,  the
   eccentricity of $s$ is the largest distance between
	  $s$ and any node of the network. Observe that the maximum
eccentricity
	  equals the diameter in the case of symmetric networks.

A typical  task in real-life radio networks
is that of performing a set of simultaneous
   and independent broadcast
operations: a {\em multi-broadcast}
operation  is to  perform $r\ge 1$
   broadcasts (from an arbitrary multiset of source nodes).
    The completion time  of a {\em Deterministic (Randomized)
  multi-Broadcast}   protocol, in short
  multi-DB (multi-RB) protocols,  is defined as follows.
A multi-DB  (multi-RB) protocol  on a radio network  has
 {\em  completion time} $t$ if, (with high probability)  every broadcast
message is received by all the nodes reachable from the source of
  the message within the first $t$ time-slots. The  termination time of a 
  multi-DB (multi-RB) protocol is defined as for DB protocols.

As for the channel bandwidth, we distinguish two models.
In the {\em Unbounded-Bandwidth} (in short UB) model \cite{CGR00},
   a node can send/receive
   messages of unbounded size
(so, a node  can send an  arbitrary large
    subset of the $r$ messages in one time-slot). 
In the  {\em Bounded-Bandwidth} (in short BB) model
   \cite{BII93},     every node can send  messages
of size at most $O(\log n + \log r)$ in one time-slot.
In this model, the completion time of the protocols    also
depends   on   the {\em congestion} (denoted as $c$) which is defined as
   the maximum number of  broadcast messages that a
   node   has
   to receive (where the maximum is computed over all possible nodes).

\subsection{Previous results}\label{sec:prev work}

   \noindent
   {\bf  Broadcast.}
   Broadcasting
in unknown radio networks has been
introduced and studied in the seminal paper \cite{BGI92} by Bar-Yehuda
{\em et al}. They proved a lower bound $\Omega(n)$ on the completion
time  of any DB protocol running on a family of   unknown, symmetric
   radio networks of diameter 3.   They also  provide a non 
spontaneous RB protocol
   having expected completion time $O((D + \log n)\log n))$.
  Thus,  the   Bar-Yehuda {\em et al}'s work  represents an important 
example in which
   randomized computations have been proved to be exponentially faster than
     deterministic ones. On the other hand, their  randomized protocol
does not terminate when no upper bound on $n$ is   known. In
\cite{ABLP89}, a lower bound $\Omega(\log^2n)$ is shown for RB
protocols that holds even for graphs of constant eccentricity (and
diameter). The best known general
    lower bound for RB protocols is
$\Omega(D\log(n/D))$, obtained in \cite{KM93}.
As for   non spontaneous DB protocols,
Bruschi and Del Pinto~\cite{BD97}  obtained a lower bound
   $\Omega(D\log n)$  for
   symmetric  networks of diameter
   $D$.   Moreover,
    an equivalent    lower
   bound   for spontaneous DB protocols on
    directed networks has been   proved by  Chlebus {\em et al}  \cite{CGGPR00}.
     The first positive results
on DB protocols in unknown radio networks
have been presented  in \cite{DKKP99}; however, this paper only
   studies   radio   networks having   restricted topologies.
In \cite{CGGPR00},   a DB protocol for
symmetric unknown networks  is presented
that has    $O(n)$  completion and termination time thus
matching the lower bound
in \cite{BGI92}, and a DB protocol   for
general  unknown  networks that  has  $O(n^{11/6})$
completion and termination  time. This protocol is the first one
that makes an explicit use of   {\em selective families}: as we will
see later, such families play a crucial role in the techniques
introduced in this paper.

  By means of    a better use of
   selective families,
     more efficient  DB protocols have been
   obtained in \cite{CGOR00}. They obtain
   a protocol having $O(n^{9/5})$  completion and termination time;   for
   networks of maximum in-degree $\Delta=O(n^a)$, they obtain
   $O(n^{1+a+H(a)+o(1)})$ completion and termination time (where $H(a)$
   denotes the binary entropy function).
They also  present a DB protocol
    having $O(n^{3/2})$ completion and termination time   and
       a DB protocol having $O(n\Delta^2\log^3n/\log (\Delta\log n))$ 
completion time
     working on unknown networks of maximum
   in-degree $\Delta$.

\noindent
The best presently known
deterministic upper bound for general unknown  networks is
$O(n\log^2n)$ and  is obtained by means of a not efficiently
constructible
DB protocol introduced by Chrobak {\em et al}
in~\cite{CGR00}.

It thus turns out  that all   previous deterministic upper bounds are
{\em superlinear} (in $n$) independently of the parameters  $D$ and 
$\Delta$ of the network. A simple analysis of such    protocols show that 
they    in fact runs  in 
superlinear time even when $D$ and 
$\Delta$ are bounded by a fixed constant.
In several wireless network applications, network topologies 
with ``small'' diameter and ``small'' degree are typically adopted in 
order to achieve fast and reliable communication \cite{AS92,L91,KR96}. It is thus natural and 
well-motivated to investigate the complexity of broadcast operations 
restricted to such topologies: the ``hope'' is to obtain protocols which 
run faster than the previous  ``general''  protocols.

\medskip
\noindent
{\bf Multi-broadcast.}
As for the UB model, Chrobak {\em et al}   \cite{CGR00} provide a
multi-DB
protocol for the {\em gossiping} operation (i.e., the special case  of $n$
simultaneous  broadcast operations, each starting from
one different node) that has $O(n^{3/2})$ completion time.
Multi-broadcast in the  BB model has been studied in~\cite{BII93}, where
     a randomized distributed
    protocol is presented that performs  the broadcast of $r$ messages
      in $O((D+r)\log \Delta\log n)$
completion time.
However, this protocol {\em does not work on
unknown networks}: it indeed assumes that every node knows its respective
neighborhood  and  that the network is symmetric.
   Moreover, it requires a set-up phase in which a Breadh First Search
    tree is computed in
$O((n+D\log n)\log \Delta)$ time-slots.

    \subsection{Our results}
\noindent
{\bf Broadcast.}
Our first contribution is    the construction of a  family of
   graphs that yields an $\Omega (n\log D)$
lower bound on the completion time
of DB protocols on unknown networks.
  The above result applies to    both
     spontaneous protocols on directed networks and   non
   spontaneous protocols on symmetric networks. The lower bound given in
     \cite{CGGPR00} and \cite{BD97} (i.e. $\Omega(D\log n)$) improves over
   the linear lower bound given in \cite{BGI92} for
   $D \ge 3$ only when $D$ is ``almost'' linear, i.e.,
   $D=\omega(n/\log n)$. Instead, our
   lower bound implies a superlinear number of time-slots for {\em any }
   $D=\omega(1)$ and, moreover, it implies that
   the $O(n\log^2n)$ deterministic upper bound given in \cite{CGR00}
    is   almost optimal when $D=\Omega(n^{\alpha})$, for
    any constant $\alpha >0$.
   We   emphasize that our lower bound also holds   when
   every node   knows $n$   and  $\Delta$.
   A simple variant of our family of graphs  allows us to get the first  lower
   bound that also depends on  $\Delta$: we indeed    provide an $\Omega
   (D\Delta\log (n/\Delta))$
   lower bound that holds  for any $ \Delta \le n/D$.
This lower bound implies that the   bound $O(n \log^{2}n)$
given in \cite{CGR00}
    is   almost optimal whenever $D\Delta=\Omega(n)$.

    \medskip
On the other hand,   we provide a new broadcast technique that yields
the first   (non constructive)  DB protocols having
   a completion-time that {\em does not} contain $n$ as  linear factor
        but only $D$ and $\Delta$. More precisely, we      obtain an
        $O(D \Delta\log(n/\Delta)\log^{1+\alpha} n)$ upper bound, 
where $\alpha$ is any
        fixed real positive constant.

   \noindent
    Our protocols are thus not efficient when $D\Delta = \omega(n\polylog n)$.
    However, by comparing them with  our $\Omega (D\Delta\log 
(n/\Delta))$ lower bound,
      we can see that these upper bounds are
      almost optimal  when   $\Delta=O(n/D)$.
      This solves  an open problem posed in \cite{CGOR00}.
     Furthermore, when $D,\Delta= O(\polylog n)$, this is an
     {\em exponential}  improvement over the superlinear (in $n$) 
completion time
     obtained by the best previously known  (not efficietly
     constructible) DB protocols \cite{CGR00}.

      Another interesting consequence lies in a new insight into the real
      gap between randomization and determinism in radio broadcasting.
     Indeed,
    rather surprisingly, if we  compare our deterministic
    upper bounds to  the $O((D+\log n)\log n)$   upper bound
     obtained by the Bar-Yehuda {\em et al}'s RD  protocol 
\cite{BGI92}, we can easily
    state that the exponential gap between
    deterministic protocols and randomized ones holds, at least from a 
theoretical
    view point, only for   radio networks  having ``large''
   maximum in-degree (i.e. when $\Delta=\Omega(n^{\beta})$,  for some 
$\beta>0$).

\noindent The following table summarize  previous and our results
for the broadcast operation on unknown networks.

\medskip

\noindent \begin{tabular}{|c||c|c|c|}
\cline{2-4}
\multicolumn{1}{c}{}\vline & \multicolumn{2}{c}{Previous
results}\vline &
Our results \\
\cline{2-3}
\multicolumn{1}{c}{}\vline & Deterministic & Randomized & Deterministic\\
\hline
\hline
Lower bound & $\Omega(D\log n)$ & $\Omega(D\log(n/D))$ &
$\Omega(D\Delta\log(n/\Delta))$ for $\Delta \leq n/D$ \\
\cline{1-4}
Upper bound & ${\rm O}(n\log^2 n)$ & ${\rm O}((D + \log n)\log n)$
& ${\rm O}(D\Delta\log(n/\Delta)\log^{1+\alpha} n)$ \\
\hline
\end{tabular}

\medskip

  \noindent
  {\bf Multi-broadcast.}
Let us  first  consider the BB model.
By combining  the trivial lower bound $\Omega(D)$ with the fact that
    a node cannot receive more than one message per time-slot, it
is easy  to derive an $\Omega(D + c)$ lower bound for the
multi-broadcast operation   for
both randomized  and deterministic   protocols   (observe that, in the UB
model, we can only get    $\Omega(D)$ since the congestion  $c$ is
always $1$). On the other hand, we are
not aware of any lower bound
  of   the form $\Omega(f(c)\cdot g(n))$ where
both $f$ and $g$ are some unbounded  functions. Such a  kind of lower
bounds  is important   since
   it   implies that a ``perfect pipeline'' protocol (i.e., a protocol
   yielding an $O(SB(D,n) + c)$ upper bound, where $SB(D,n)$  is the best
     upper bound available for the   broadcast operation) is
not achievable.
   We  provide  the first lower bound of the kind
     defined above, even    under very restrictive topology conditions.
          We indeed derive a family
         of   graphs with  $\Delta=2$ that forces any
         multi-DB protocol to perform at least
      $\Omega(c + (\frac{c}{\log c}+D) \log n)$ time-slots to complete
        multi-broadcast operations.
Then, we derive an  $\Omega(c+\frac{c}{\log c}\log n +D \log \frac nD)$
lower bound    to  multi-RD  protocols.
Hence,  perfect pipelining is not achievable even with the
help of distributed random choices.

\noindent
We observe that the above  lower bounds also  hold in  presence of collision
detection   and when the nodes  know   $n$  and/or  $\Delta$.

On the other hand,  we combine  a variant of
   our  (single) broadcast technique
     with a suitable ``local'' scheduling   (that  solves
    the  congestions arising inside every node) in order to get
a  multi-DB protocol    for the BB
model. This protocol has $O((D+c) \Delta^2  \log^{2+\alpha} n)$ completion
time (where $\alpha$ is any
        fixed real positive constant) and it can be converted into an
        efficiently constructible one having $O((D+c) \Delta^2 
\log^{3+\alpha} n)$ completon time.

\noindent
By comparing the above upper bounds with our deterministic lower bound,
we have that our multi-DB protocols
turn out to be ``almost'' optimal (i.e., only a polylogarithmic factor away
from the lower bound)  when $\Delta=O(\polylog n)$. We also emphasize
that, for     $\Delta =O(1)$, our
deterministic upper bound   is almost
equivalent to the $O((D+r)\log \Delta\log n)$ randomized upper 
bound~\cite{BII93}
  in which it is
even assumed that the network is symmetric and the nodes know their respective
neighborhood.

As for the UB model, since arbitrary large concatenation of the messages
inside a node can be sent along the outgoing edges, we can use a simpler
version of our multi-DB  protocols. In this
version,
the local scheduling is not required and, thus, we get an
$O( D \Delta^2  \log^{2+\alpha} n)$    upper bound.

   \subsection{Organization of the paper}
   
     Section \ref{sec:connection} provides  an overview of the  
     connections between the issue of  radio broadcasting
     and some  combinatorial 
   concepts and results. In Section \ref{sec::comby}, the proofs of such 
   combinatorial results are given. Section \ref{sec:single} describes   the 
   results on the  broadcast operation.
    In Section \ref{sec::multy}, the results
   on the multi-broadcast operation are presented.
   Finally, Section \ref{sec::CONCLUSIONS} discusses the obtained results and proposes
   some open problems.

     \section{Adopted techniques and some new 
     combinatorial  results: an overview}
\label{sec:connection}

The proofs of our upper and  lower bounds   exploit   some
    new combinatorial results that we believe to  have a {\em per se} interest.
In this section, we provide a description of such results and we
outline their connection with broadcast and multi-broadcast operations.

In  \cite{CGGPR00}, Chlebus {\em et al} introduced the use of {\em
selective families} in designing DB protocols in unknown  networks.
In what follows, the set      $\{ 1, \ldots , n\}$ is denoted as $[n]$.

\begin{defi}\label{def:sel-fam}
Let $n$ and $k$ be any integers with
$k \leq n$. A family $\sF$ of subsets of $[n]$ is
{\em $(n,k)$-selective} if, for every non empty
subset $Z$ of $[n]$ such that $ |Z| \leq
k$,
there is a set $F$ in $\sF$ such that $ |Z \cap F| =1$.
\end{defi}

\noindent
In fact, selective families  can  be used to design {\em oblivious} DB
protocols. A protocol is oblivious when its actions can be
scheduled in advance. Let $\sF=\{F_1,\ldots,F_m\}$ be an
  $(n,\Delta)$-selective family. Then, it is easy
to define an oblivious DB protocol on a network of $n$ nodes and
maximum in-degree
$\Delta$ (this protocol has been introduced in \cite{CGGPR00,CGOR00}).

\begin{verse}
A node $u$ transmits
at time-slot $t\leq m$ iff it has received the source message (i.e. it
is informed) and $u \in F_t$.
\end{verse}
  During the execution of these $m$
time-slots, thanks to the  selective property of $\sF$, for each
set of $d\leq \Delta$ nodes, there is at least one time-slot in which only one
of the nodes of the set can transmit. This guarantees that at least 
one non informed
node gets informed. By iterating this process $n$ times,
the broadcast is completed.
Thus the completion time of such DB protocol
is    $n|\sF|$.

  We prove that there
  exist $(n,k)$-selective families of size $O(k \log (n/k))$. We also
  prove that this upper bound is optimal, that is, any $(n,k)$-selective
  family has size $\Omega(k \log (n/k))$.  On one hand, such small
  selective families are combined with a new broadcast technique in
  order  to obtain our DB protocols. On the other hand, the
  lower bound on the size of selective families allows us to obtain
  lower bounds  on the
  completion time of DB protocols.

In designing multi-DB protocols, we introduce families of sets having  a
stronger selective property.
   \begin{defi}\label{def:stron-sel-fam}
Let $k \leq n$.  A family $\sF$ of subsets of $[n]$ is
{\em $(n,k)$-strongly-selective} if for every
subset $Z$ of $[n]$ such that $
|Z| \leq k$ and for every element
   $z\in Z$ there is a set $F$ in $\sF$ such that $ Z
\cap F =\{ z\}$.
\end{defi}

\noindent
If we have at hand an $(n, \Delta +1)$-strongly-selective family
$\sF=\{F_1,\ldots,F_m\}$, then it is easy to define an oblivious
  multi-DB protocol for the UB model on a network of $n$ nodes and maximum
in-degree $\Delta$. A node $u$ transmits (all the messages it knows)
at time-slot $t \leq m$ iff $ u \in F_t$. By the strongly-selective
property of     $\sF$, for each  set $X$ of
$d\leq \Delta +1$ nodes and for each node $u\in X$,
there is at least one time-slot in which only $u$
transmits among the nodes in $X$. This guarantees that every message
reaches at least a new node during the $m$ time-slots. By iterating this
process $n$ times  the multi-broadcast is completed in $n|\sF|$ time-slots.

  \noindent
  Similarly to our single DB protocol,
  our multibroadcast protocol exploits the existence and the construction of
  strongly-selective families of small size. Actually, 
strongly-selective families are
  a new appearance of a well-known notion.
   Let $\sF=\{F_1,\ldots,F_m\}$ be a
   family of subsets of $[n]$ and consider the matrix $M^{\sF}$ of $m$ 
rows and $n$ columns
   where $M^{\sF}_{t,i}$ is set to 1 iff $i\in F_t$. It turns out that
   $\sF$ is an $(n, k)$-strongly-selective  family iff the
    {\sc or} of any set of at most $k$ columns of $M^{\sF}$ covers
  only the
  columns of the set. Hence the columns of $M^{\sF}$  form a
  {\em superimposed code} \cite{KS64}.
  Superimposed codes are also known in
  combinatorics as {\em cover free families\/} \cite{EFF}. Our
  protocol uses the $(n,k)$-strongly-selective families of size
  $O(\min\{n, k^2\log n\})$ whose existence is proved in \cite{EFF}.
This existence proof does not provide an efficient construction.
However, efficient construction of such families can be found in
\cite{KS64}: in this case the size is  $O(\min\{n, k^2\log^{2} n\})$.

\noindent  Since a lower bound
  on the size of strongly-selective
  families determines a lower bound on the completion time of
  {\em our} multi-broadcast technique,
  we have also investigated this
  combinatorial aspect.
In \cite{STOC96}, Chaudhuri and Radhakrishnan  obtain a  lower bound
     $\Omega((k^2/\log k)  \log n)$ for sufficiently large $k$ such
  $k\le  n^{1/3}$. Our contribution here is the extension to every
  $k$ of that bound, that is, $\Omega(\min \{n, (k^2/\log k)  \log n\})$.
This      implies
     that there are no significantly smaller strongly-selective
     families than those  adopted by our protocols.

\noindent
  As  in the case of selective families, we tried to exploit
  strongly-selective families to obtain good lower bounds on the
  completion time of multi-DB protocols. But we could not obtain
  anything better than the lower bound for single broadcast.

In the BB model, besides the interference problems,
the  bound on the channel bandwidth
yields further delays due to the congestion inside  the  nodes of the
network. We exploit this aspect to obtain a non trivial lower bound.
Consider the following situation that can happen during the execution
of a multi-broadcast protocol. There are $n$ nodes each of them having
$r$ messages. They are not connected each other, and exactly two of
them are the only in-neighbours of another node $u$. Thus, all the $r$
messages can be received by $u$ only if, for every pair of nodes and
for each message, there is a time-slot in which exactly one of the
nodes of the pair transmits that message and the other does not
transmit anything. This suggests to introduce
     the following notion.

  \begin{defi}
Two sequences $\vec x$ and $\vec y$ of equal length over  the alphabet
$\{0\} \cup [r]$ are $r$-different if for any $z\in [r]$ there is a
coordinate $i$ for which $\{x_{i},y_{i}\}=\{z,0\}$.
\end{defi}

\noindent
 From the situation described above, we will show that
a protocol which    performs any multi-broadcast operation on
unknown networks within $t$ time-slots must yield a set of $n$
pairwise $r$-different sequences of length not greater than $t$.   By combining
this connection with a new lower bound on the length of such
sequences, i.e. $\Omega((r/\log r) \log n)$, we derive the lower bounds
for the multi-broadcast operation on the BB model.

\section{Combinatorial results}\label{sec::comby}

    \subsection{Selective families} \label{sec:sel-fami}

\noindent
We  now   show the existence of $(n,k)$-selective families of small size
by a suitable application of the probabilistic
method~\cite{AS92}.
  \begin{thm}\label{th::upperbound-sel}
For any $n>2$ and $k\geq 2$, there exists ($n,k$)-selective family of size
$O(k\log( n/k))$.
\end{thm}

\proof
In the sequel we say that a family $\sF$ is selective for a
family $\sS$ if, for each $S \in \sS$, there is a set $F\in \sF$ such
that $| F\cap S| =1$.
Let $\sS_i$, $1\leq i \leq \lceil \log k
\rceil$, be the family of all the subsets
of $[n]$ having size in the range $(2^{i-1},2^i]$. Consider now a
family $\sF_i$  of $l_i$ sets (the value of $l_i$ is specified
later)  in which each set is defined by randomly picking   every element
of $[n]$ independently,  with probability $\frac{1}{2^i}$.

\noindent Fix a set $S\in\sS_i$ and   consider a  set $F\in \sF_i$;
  then it holds that

$$
\mbox{ Pr}[|F\cap
S|=1]=\frac{|S|}{2^i}\left(1-\frac{1}{2^i}\right)^{|S|-1} >
\frac{|S|}{2^i}\left(1-\frac{1}{2^i}\right)^{2^i}\geq \frac{|S|}{4\cdot
2^i}\geq
\frac{1}{8}
$$
where the second inequality is due to the fact that
$\left(1-\frac{1}{t}\right)^t \geq \frac{1}{4}$ for $t\geq 2$.

\noindent The sets in $\sF_i$ have been constructed independently, so, from
the above inequality, the probability that $\sF_i$ does not select   $S$ is
  at most
$$ \left(1-\frac{1}{8}\right)^{l_i}\leq e^{-\frac{l_i}{8}}$$.

\noindent Hence we have that
\[ \mbox{Pr[$\sF_i$ is not  selective for  $\sS_i$]}\leq \sum_{S\in
\sS_i}\mbox{Pr[$\sF_i$ doesn't select $S$]}\leq
\sum_{d=2^{i-1}+1}^{2^i}\Cnk{n}{d} e^{-\frac{l_i}{8}}\]

\noindent
By choosing $l_i>8\ln\left(\Cnk{n}{2^{i-1}}2^i\right)$, we get
$$\sum_{d=2^{i-1}+1}^{2^i}\Cnk{n}{d} e^{-\frac{l_i}{8}} \leq
\sum_{d=2^{i-1}+1}^{2^i}\frac{\Cnk{n}{d} }{\Cnk{n}{2^{i-1}} 2^i} \leq
\frac{2^{i-1}}{2^{i}}=\frac{1}{2}$$
Since  $\log \Cnk{n}{t} = O(t\log (n/t))$, it holds that $l_i =
O(2^i\log (n/2^i))$ thus there exists a family
$\sF_i$ selective for $\sS_i$ and having size $O(2^i\log (n/2^i))$.
Finally, we consider the $(n,k)$-selective
family
$$\sF=\bigcup_{i=1}^{\lceil \log k \rceil}\sF_i$$

\noindent whose  size is
$$\sum_{i=1}^{\lceil \log k \rceil}O(2^i\log(n/2^i))=O(k\log (n/k)).$$

\eproof

In what follows, we provide a lower bound on the size of selective families.
To this aim, we
  make use of     the notion of {\em intersection free} family

\begin{defi}
Let $l \leq k \leq n$. A family $\sF$ of $k$-subsets of $[n]$ is
   {\em $(n,k,l)$-intersection free} if $| F_1 \cap F_2|\neq l$ for every
   $F_1$ and $F_2$ from $\sF$.
\end{defi}

\noindent
Roughly speaking, the intersection free property is somewhat
``complementary'' to the selectivity property we are using in this
paper. So, even though an explicit
mathematical connection between the two properties
   will be determined  later, the
reader can already imagine our interest in introducing the following
result obtained by
   Frankl and F\"uredi.

\begin{thm}\label{thm:furedi}{\em \cite{FF}}
Let $\sF$ be an  ($n,k,l$)-intersection free family where $2l+1\geq k$
and $k-l$ is a prime power. Then it holds that

$$|\sF| \leq \Cnk{n}{l}\Cnk{2k-l-1}{k}\left/\Cnk{2k-l-1}{l} \right.$$
\end{thm}

\noindent
In particular, we  first prove        the following
consequence of the above theorem

\begin{cor} \label{cor:furedi}
Let $\sF$ be an  ($n,k,\frac{k}{2}$)-intersection free family where
$k$ is a power of $2$ and $k\leq \frac{n}{64}$. Then  it holds that

\[ \log |\sF| \leq \frac{11k}{12}\log \frac{n}{k} \]
\end{cor}

\proof
By using the  following inequalities   involving binomial coefficients

\[  \left(\frac{a}{b}\right)^b\leq \Cnk{a}{b}\leq
\left(\frac{ea}{b}\right)^b  , \ \
   \Cnk{a-1}{b}=\frac{a-b}{a}\Cnk{a}{b} \]

\noindent
we obtain

$$\log |\sF|\leq \log
\left(\Cnk{n}{k/2}\frac{\Cnk{3k/2-1}{k}}{\Cnk{3k/2-1}{k/2}} \right) =
\log \left(
\frac{1}{2}\Cnk{n}{k/2}\frac{\Cnk{3k/2}{k}}{\Cnk{3k/2}{k/2}}\right) \leq $$

$$\leq \log \left( \frac{1}{2}
\left(\frac{2en}{k}\right)^{k/2}\left(\frac{3e}{2}\right)^k 3^{-k/2}
\right) =$$

$$\frac{k}{2}\log \frac{n}{k} + \frac{k}{2}\log 3 +\frac{3k}{2}\log e
-\frac{k}{2} -1 < \frac{k}{2}\log \frac{n}{k} + \frac{5}{2}k  <
\frac{11k}{12}\log {\frac{n}{k}}$$

   \eproof

\noindent
We are now ready to prove the lower bound.

  \begin{thm}\label{thm:nostro}
For any $n>2$, let $\sF$ be an  ($n,k$)-selective family
with $2 \le k\leq \frac{n}{64}$.
Then it holds that
\[ |\sF| \geq \frac{k}{24}\log\frac{n}{k} \]
\end{thm}

\proof
Let $k'$, $\frac{k}{2} <  k' \leq k$, be a power of 2. Let $\chi(G)$ be the
chromatic number of  the graph $G$ whose vertices are all the $k'$-subsets
of $[n]$
and whose edges connect vertices having exactly $\frac{k'}{2}$ common
elements. The theorem is an immediate consequence of
the following inequalities

\begin{equation}\label{eq::prima}
\log \chi(G) \ \ge \  \frac{k}{24}\log\frac{n}{k}
\end{equation}

  \begin{equation}\label{eq::secon}
    |\sF | \ge    \log \chi(G) .
    \end{equation}

\noindent
We first prove Ineq. \ref{eq::prima}. For any graph $G(V,E)$ with 
stability number
$\alpha(G)$ it holds that

\begin{equation}\label{eq::third}
  \chi(G)\geq \frac{|V|}{\alpha(G)} .
   \end{equation}

  Clearly a stable set of vertices in  $G$ forms an
($n,k',\frac{k'}{2}$)-intersection free family satisfying the conditions of
Corollary~\ref{cor:furedi}. hence,  from Ineq. \ref{eq::third} and
Corollary ~\ref{cor:furedi}, we have that

$$\log \chi(G)\geq \log |V| - \log \alpha(G) \geq \log
\Cnk{n}{k'}-\frac{11k'}{12}\log \frac{n}{k'}\geq $$

$$\geq k'\log\frac{n}{k'}-\frac{11k'}{12}\log \frac{n}{k'} =
\frac{k'}{12}  \log \frac{n}{k'}\geq \frac{k}{24} \log \frac{n}{k} . $$

\noindent We now  prove Ineq. \ref{eq::secon}.
Here we use the straightforward inequality
\begin{equation}\label{eq::B}
  \chi(\bigcup_{i=1}^t G_i)\leq \prod_{i=1}^t\chi(G_i)
  \end{equation}

  \noindent
  that holds for any set
of graphs having the same set of vertices.\\
Let be $\sF=\{F_1,\cdots,F_{|\sF|}\}$. We define  the graph $G_i$, $1\leq i
\leq |\sF|$, by setting $V(G_i)=V(G)$ and by drawing an edge between two
vertices of $G_i$ if they are adjacent in $G$ and furthermore $|F_i \cap
X|=1$, where $X$ is the
symmetric difference of the sets corresponding to the two vertices. Since
$\sF$ is a  ($n,k$)-selective family and the symmetric difference of these
sets has cardinality $k'$, for any edge of $G$, there will be at least a
graph $G_i$ having this edge.   Hence we have $G=\cup_{i=1}^{|\sF|}G_i$.
It thus follows that
$$\log \chi(G)=\log \chi(\bigcup_{i=1}^{|\sF|} G_i)\leq \log 
\prod_{i=1}^{|\sF|}\chi(G_i)
= \sum_{i=1}^{|\sF|}\log \chi(G_i)\leq |\sF|$$
where the first inequality follows from Ineq. \ref{eq::B} and
the  last inequality follows by noting that the graphs $G_i$ are
bipartite graphs (i.e. $\chi(G_i)\leq 2$): indeed, for any  two
adjacent  vertices in $G_i$ one has odd intersection with the 
elements of $F_i$ and
the other has even intersection.
\eproof

\subsection{Strongly-selective families }

\noindent
    In \cite{DR},  Dyachkov  and Rykov proved a lower bound $\Omega(c_k\log
   n)$ on the   size of $(n,k)$-strongly-selective families, where
    $c_k\rightarrow \Theta(k^2/\log k)$  for  $k\rightarrow \infty$.
    Observe  that this {\em does not} imply the standard two-variable
    lower bound
   $\Omega((k^2/\log k)\log  n)$: for instance when $k=\Omega(n)$,
   this would imply a
   lower bound  $\Omega(n^2)$. The latter   is clearly false. Indeed,
   the family    consisting of all the singletons from $[n]$ is
   $(n,k)$-strongly-selective for any $k=1,\ldots , n$, and it has size
   $n$. In \cite{STOC96}, Chaudhuri and Radhakrishnan  obtain a  lower bound
     $(k^2\log n)/(100\log k)$ for sufficiently large $k$ such
     that\footnote{Notice that the conditions on $k$ are stated immediately
     before Lemma 5.1 of  \cite{STOC96}} $k\le
     n^{1/3}$.
     Our   contribution here is     the  generalization (and an
     improvement)
      of the   Chaudhuri and Radhakrishnan's result.

We prove a lower bound that is only an $O(\log k)$ factor away from
the $O(\min\{n, k^2\log n\})$    bound in \cite{EFF}.

   \medskip
\noindent
\begin{thm}\label{th:strongly-low}
Let $\sF$  be an  $(n,k)$-strongly-selective family.
\begin{description}

\item[{\bf i).}] If
   $3\leq k\leq \sqrt{2n}-1$  then it  holds that
$ |\sF| \geq \frac{k^2}{48\log k}\log n$.

   \item[{\bf ii).}] If $k\ge \sqrt{2n} $
then it holds that $ |\sF| \geq n$.
\end{description}
\end{thm}

\noindent
\proof
{\bf i).} The proof  relies on  a result by F\"uredi \cite{F}
       and a result by Bassalygo  \cite{DR} on superimposed
       codes.  For the sake of convenience, we state such results in
       terms of strongly-selective families.

Let  $\sF$
    be an	 $(n,k)$-strongly-selective family  then Bassalygo proved that

\begin{equation}\label{Eq::uno}
    |\sF|\geq \min\left\{ \Cnk{k +1}{2},n \right\} ,
    \end{equation}
\noindent
and F\"uredi proved that, for $k\ge 3$,
\begin{equation}\label{Eq::due}
     n \leq k -1 + \Cnk{|\sF|}
{\left\lceil
\frac{|\sF| -k +1}{\Cnk{k}{2}}
\right\rceil}
\end{equation}

\noindent
Let $3\le k\le \sqrt{2n}-1$.
  From Ineq.~\ref{Eq::due} and   the   inequality
   $\Cnk{a}{b}\leq \left(\frac{ea}{b}\right)^b$, we get

   \[\frac{k(k-1)}{\log\frac{e|\sF|k(k-1)}{2(|\sF|-k+1)}}\log(n-k+1)\leq
   2|\sF|-3k +2 +k^2\]
Since  $\frac{k^2}{2}\leq k(k-1)<k^2$ and  $\sqrt{n}< n-k +1$, it follows
that

$$
\frac{k^2}{4\log\frac{e|\sF|k^2}{2(|\sF|-k+1)}}\log n\leq
   2|\sF|-3k +2 +k^2
$$

   \noindent
   Moreover, since   $k\leq \sqrt{2n}-1$,    Ineq.~\ref{Eq::uno} implies that
   \[
   |\sF|\geq \frac{k^2 +k}{2}\]
    and then
   \[ \frac{|\sF|}{2(|\sF|-k+1)}\leq 1   \ \mbox{ and } \
      -3k +2 +k^2 \leq 2|\sF| \]
We  thus    obtain

   \[\frac{k^2}{4\log(e k^2)}\log n\leq
   4|\sF| . \] Finally, since $k \ge e$, $ |\sF| \geq 
\frac{k^2}{48\log k}\log n$.

  \noindent
  {\bf ii).}  When $k\ge \sqrt{2n}$, the thesis follows immediately
    from Ineq. \ref{Eq::uno}.
   \eproof
%
%
%

    \subsection{Sets of pairwise r-different 
sequences}\label{app::prf-of-lm r-good}

\noindent Our goal in this section is to prove a lower bound on the
length of $n$ sequences which are pairwise $r$-different.
In the sequel,  we will make  use of the
binary entropy function $h(t)=-t\log t-(1-t)\log (1-t)$ .

\noindent
The proof of our lower bound relies on  the
  following nice theorem proved in \cite{EJ}

\begin{thm}\label{FK}{\em \cite{EJ}}
Let $S$ be a subset of  ${ [r] \choose 2}$ and $C$ be a set of
sequences
of   length $m$ over  the alphabet $[r]$ with the property that for each
   $\{x,y\} \in {C \choose 2}$ and
   $\{a,b\}\in S$ there exists an    $i \in [m]$ such that $\{x_i,y_i\} =
\{a,b\}$. Then it holds that
$$\log |C| \leq m\max_P \min_{\{a,b\}\in
S}\left\{(p_a+p_b)h\left(\frac{p_b}{p_a +p_b}\right)\right\}$$
where, in the maximum, $P$ is running over all the probability
distributions on $[r]$.
\end{thm}

\begin{thm}~\label{intro:lem:r-diff}
Let $M(n,r)$ denote the minimum length  of $n$ sequences which are
  pairwise $r$-different. Then
\[ M(n,r)  =  \Omega \left (\frac{r}{\log r}\log n \right).\]
    \end{thm}
\proof
   Let $C$ be a set    of  $n$     sequences which are pairwise
   $r$-different and
       define
$S=\left\{ \{0,i\}|i\in [r]\right\}$. From Theorem~\ref{FK} we have that:
\[
\log n \leq M(n,r)\max_P \min_{i\in
[r]}\left\{(p_0+p_i)h\left(\frac{p_i}{p_0 +p_i}\right)\right\} .
\]
\noindent
   Let
   $ f(x)=(p_0+x)h\left(\frac{x}{p_0 +x}\right)$, $0\leq x \leq 1$. We have
    that $f'(x)=\log\frac{p_0 +x}{x}\geq 0  $ for any  $0\leq x \leq 1$. Thus
$f$
    is  not decreasing and     we can restrict the
search of the maximum value, in the right hand of the above inequality,
to those probability
distributions in which $p_i$ have  the same value
for all $i\in [r]$, i.e., $p_i=(1-p_0)/r$.

\noindent
We thus consider for any $x \in[0,1]$,    the probability distribution
$p_i=\frac{x}{r}$ for  $i\in [r]$ and $p_0=1-x$. Then the inequality can
be written as:
\[\log n \leq M(n,r)\max_{x \in[0,1]}
\left\{\left(1-x+\frac{x}{r}\right)h\left(\frac{\frac{x}{r}}{1-x+\frac{x}{r}}\right)
\right\} . \]

\noindent
In order to prove the theorem, we  show that
\[ \max_{x \in[0,1]}
\left\{\left(1-x+\frac{x}{r}\right)h\left(\frac{\frac{x}{r}}{1-x+\frac{x}{r}}\right)
\right\}
  =  O\left( \frac{\log r}{ r}\right) . \]
Indeed,   the function
\[f(x)=
\left(1-x+\frac{x}{r}\right)h\left(\frac{\frac{x}{r}}{1-x+\frac{x}{r}}\right)
\]
   can be written  as
   \[ f(x)=(1-x)\log \left(1 + \frac{x}{r(1-x)}\right)
+\frac{x}{r} \log \frac{r-rx +x}{x} \ .\]
Then,  by using the well known inequality $1 +t \leq e^t$ (that holds
for any real  $t$), we get:

$$ (1-x)\log \left(1 + \frac{x}{r(1-x)}\right) \leq  (1-x)\log 
e^{\frac{x}{r(1-x)}}
\leq  \frac{x}{r} \log e \leq  \frac{1}{r} \log e  =  O\left(
\frac{1}{r}\right) . $$
\noindent
It thus suffices  to prove that

\[ g(x)=\frac{x}{r} \log \frac{r-rx +x}{x} =  O\left(\frac{\log
r}{r}\right) . \]

\noindent
Since   $\frac{r-rx +x}{x}$ is a decreasing function in the interval set
$[\frac{r}{3r-1},1]$ then, for $x\in [\frac{r}{3r-1},1]$,

\[g(x)\;\leq\; \frac{x}{r}\log 2r\;\leq\; \frac{1}{r}\log 2r \in
O\left(\frac{\log
r}{r}\right) \]

\noindent Furthermore,

\[ g'(x)=\frac{1}{r}\log \frac{r-rx
+x}{x}- \frac{1}{(r-rx +x)\ln 2}  \]  is strictly positive in the interval set
$[0,\frac{r}{3r-1}]$. Thus,  in $[0,\frac{r}{3r-1}]$,  the function  $g(x)$
is increasing.
Hence, for $x\in [0,\frac{r}{3r-1}]$,

\[ g(x)\leq g\left(\frac{r}{3r-1}\right)=\frac{1}{3r-1}\log 2r =
O\left(\frac{\log
r}{r}\right) \]
\noindent This completes the proof.
\eproof

\section{Broadcast operations}\label{sec:single}

\subsection{The lower bounds}

In this section, we show the existence of an infinite family of directed
graphs that force any DB protocol to perform, in the worst-case,
   $\Omega(n\log D)$ time-slots before completing  a broadcast.
Then, we provide a simple variant of this family of graphs yielding a
lower bound that also depends on $\Delta$.
Our lower bound holds for   the UB model (and, thus, for the BB model
too).
We first formalize the notion of DB protocol according to  \cite{BGI92}.

   \begin {defi}\label{def::the single model}
    A {\em Deterministic distributed Broadcast} DB protocol
    $P$ is a   protocol that works in time-slots
    (numbered $0,1,\ldots $) according to the following rules.

    \begin{enumerate}
    	\item  In the initial time-slot a specified node (i.e. the {\em
    	source}) transmits a message (called the {\em source  message}).

    	\item  In each time-slot, each node either acts as   transmitter
    	or as   receiver or is non active.

    	\item  A node receives a message in a time-slot if and only if
    	it acts as  receiver   and {\em exactly} one of its
    	$in$-neighbors acts as  transmitter in that time-slot.

    	\item  The action of a node in a specific time-slot is
    	 a function of   its own label, the
    	number of the current time-slot $t$, and the messages received
    	during  the previous time-slots.

    \end{enumerate}

    \end {defi}

   \begin{thm}\label{thm:LOWER B}
For any DB protocol $P$,  for any  $n$
and for any $   D \leq   n/6$, there exists an $n$-node directed
graph $G^{P}$ of maximum eccentricity $D$
such that $P$ completes broadcasting on $G^P$ in   $\Omega (n\log D)$
time-slots. The lower bound holds even when every node knows $n$.
\end{thm}

\proof
   The graph $G^{P}$ is a layered $n$-node graph with $D+1$ levels
$L_0, L_1, \ldots ,
   L_{D}$; Level $L_0$ contains only the source $s$,   level $L_j$
   has  no more than $\lfloor n/ (2D) \rfloor$ nodes for $j=1,\ldots,D-1$
and, finally, the
     level $L_D$ contains  all the   remaining nodes. All nodes of
     $L_{j-1}$ have   outgoing edges to all nodes in  $L_{j}$. As we will
     see later, the actions specified by $P$ determine the node
assignment
      in the levels $j\ge 1$ in such a way that the protocol is
forced
      to execute $\Omega((n/D)\log D)$ time-slots in order to successfully
      transmit the source message between
       two consecutive levels. This
      assignment will be performed by induction on the levels.

      \noindent
   From Theorem~\ref{thm:nostro},
       there exists a constant $c>0$ such that, if $2\le D \le n/6$,
     any     $(\lceil n/2\rceil , \lfloor n/(2D) \rfloor)$-selective family
must have size at least
     $T$, where
$ T = \lfloor { c \frac nD \log D } \rfloor$.

   \noindent
    The  theorem is then an easy consequence  of  the following

     \begin{verse}
      {\bf Claim 1.} {\em For any $j = 0, \ldots ,  D -1 $, it is 
possible to assign
          nodes in $L_0, L_1, \ldots , L_j$
      in such a way that $P$ does not broadcast the
      source  message to level $L_{j}$ before the time-slot $j \cdot T$.}

\proof
    The proof is by induction on $j$. For $j=0$,
     the claim is trivial. We thus assume the  thesis be true for
     any   $j$ and we prove it for $j+1$. Let us define

     \[ R = \{ \mbox{nodes     not already assigned to levels }
     L_0,\ldots , L_{j} \}
     \]
     Notice that $|R| \ge \lceil n/2 \rceil$. In fact

     \[ |R| = n - \sum_{h=0}^{j} |L_h|
       \ \ge \ n -
     \left(\left\lfloor \frac {n}{2D}\right\rfloor \right) (D-2) -1\
     \ge \
     \left\lceil \frac n2 \right\rceil \]

      \noindent
      Let $L$ be an arbitrary subset of  $R$. Consider the following
      two cases:
      $i)$ $L_{j+1}$ is chosen as $L$, and $ii)$ $L_{j+1}$ is chosen 
as $R$ (i.e.
      all the remaining nodes are assigned to $L_{j+1}$).
      In both
cases, the predecessor\footnote{Given a graph $G$, the predecessor 
subgraph $G_u$ of a node
$u$ is the subgraph of $G$ induced by all nodes $v$ for which there
exists a
directed path from $v$ to $u$.} subgraph $G^{P}_u$ of any node $u \in 
L$ is that induced
by $L_0 \cup L_1 \cup
    \ldots L_{j+1} \cup \{u\}$ in $G^{P}$. It follows
    that the behaviour of node $u$,
    according to protocol $P$, is the same in both cases.
We can thus consider the behavior of $P$ when $L_{j+1} = R$. Then, we define

    \[ F_t = \{ u\in R \  | \ u \mbox{ acts as
    transmitter at time-slot } j \cdot T+t \} .
   \]
    and   the family $ \sF = \{ F_1,\ldots , F_{T-1} \}$
     of subsets from $R$.
  Since $| \sF| < T$,   $\sF$ is not
    $(\lceil n/2\rceil, \lfloor n/(2D) \rfloor )$-selective; so, a subset
$L \subset R$ exists
    such that $|L| \le \lfloor n/(2D) \rfloor$ and
     $L$ is not  selected by $\sF$ (and thus by  $P$)
     in any time-slot $t$ such that  $jT + 1\leq t \le (j+1)T -1$.
     The proof is completed by choosing $L_{j+1}$ as $L$.
      \end{verse}

\eproof

   \medskip
   \noindent
   \begin{thm}\label{cor::LOW.BOUND-d}
Let    $P$ be a    DB protocol. Then, for  any  $n$,
     for any  $D \leq   n/6$, and for any $\Delta \le n/D$,
   there exists an $n$-node directed
graph $G^{P}$ of maximum eccentricity $D$ and   in-degree bounded by
$\Delta$
such
that $P $ completes broadcasting on $G^{P}$ in   $\Omega (D\Delta\log
(n/\Delta))$
time-slots. The lower bound holds even when every node knows $n$
and $\Delta$.
   \end{thm}

   \proof
   The proof is based on the same construction of   the proof
    of Theorem ~\ref{thm:LOWER B}.
   The only difference is that, for every  $j=1,2,\ldots D-1$,
      level $L_{j}$
    of $G^{P}$ consists  of at most $\Delta$ nodes   and   $L_{D}$
    (consisting of all the remaining nodes) is  connected
    to the previous level in such a way that
    the maximum in-degree is kept not larger than  $\Delta$.
\eproof

\subsection{The upper bounds}\label{upp_b}
This section provides a DB protocol for unknown networks. For case of
exposition, we first describe the protocol that assume the knowledge
of $n$ and $\Delta$. Then, we show how to extend the same technique
to the cases in which $\Delta$ and $n$ are not known by the nodes.


\noindent
 The following   protocol  assumes the knowledge of  
$\Delta$ and   $n$.

\medskip
\noindent
{\bf Description of Protocol $\prota (n,\Delta)$.}
   The protocol  uses an    $(n,\Delta)$-selective family $\sF$.
It starts by setting all the nodes to the active state, and by
let $s$ transmit the source message.     After the first time-slot,  it
   turns $s$ to the non active state.
   Then, it performs     a sequence of  consecutive identical {\em phases}.
    Let us fix an arbitrary
ordering for   the sets of  $\sF$;  at time-slot $j$ of    {\em phase} $i$,
each  node $v$ acts according to the following rule: $v$  transmits
the source message  along its outgoing edges if and only if

\begin{itemize}

\item[1)] the label of $v$ belongs to the $j$-th set of $\sF$, and

\item[2)] $v$ has received the source message {\em for the first time}
  during the  phase $i-1$.

   \end{itemize}

\noindent
   After the phase in which a node $v$ acts as a transmitter, it
   turns to the non active state   (so, this is a first  change w.r.t. the
    straightforward
     protocol described in Section~\ref{sec:connection}).
The active  nodes that, at at time-slot $j$ of any phase, have a state
     not satisfying   Conditions  $(1)$ and $(2)$ act as receivers.
      Observe that  Condition $(2)$ is the
     {\em key} difference between our technique and the straightforward
     one. As we will see in the analysis of
     the protocol, this difference  will
    play a crucial role in
order to achieve an upper bound not containing
the  linear   factor    $n$.

     \begin{thm}\label{single::upper bound}
    Protocol  $\prota(n,\Delta)$  completes broadcasting {\em and terminates}
      in     $O(D \Delta\log (n/\Delta))$ time-slots on  any $n$-node
graph of maximum eccentricity $D$ and  maximum in-degree     $\Delta$.
\end{thm}

\proof
Since, Theorem~\ref{th::upperbound-sel} implies that $|\sF|=
O(D \Delta\log (n/\Delta))$,
the  thesis is an easy consequence of the following claim.

\begin{verse}
{\bf Claim.}
A node  $v$  receives (for the first time)
the source  message at phase $i$ of protocol $\prota(n,\Delta)$
     if and only if   $v$ is at distance $i+1$
   from the source $s$.

   \proof
   The proof is by induction on $i$.
($\Leftarrow$). For $i=0$ the Claim is obvious.
We thus assume that all nodes at distance $i$ have received the source
message during  phase $i-1$. Let us consider a  node $v$ at distance
$i+1$ during phase $i$.
This node has at least one informed in-neighbor at distance $i$.
According to the protocol,
only the   neighbors of $v$  informed in phase $i-1$
   will act as transmitters in phase $i$. Then, from the 
$(n,\Delta)$-selectivity
   of    $\sF$, there will be a step of phase $i$, in which only one
   of these informed  in-neighbors will transmit to $v$.
   ($\Rightarrow$). If $v$ is not at distance $i+1$ from $s$,
    then two cases may arise. If
   $v$ is at distance less than $i+1$ then, by the inductive
   hypothesis, $v$
   has been informed before phase $i$. Otherwise, $v$ is at distance 
greater than
$i+1$, so none of its in-neighbors has been informed before phase $i$.
\end{verse}

\eproof

\noindent
The next protocol assumes   the knowledge of $n$.

\medskip

\noindent
{\bf Description of Protocol $\protb(n)$.}
Each node runs a sequence of  {\em phases}, each of them
   consisting of
$\lceil \log n \rceil$ time-slots. In time-slot $l$ ($1\le l\le \lceil\log
n \rceil$)
of phase $h$,
each node runs time-slot $h$ of   $\prota(n,2^l)$. Furthermore,
   if a node $v$
is set to the non active state in   a time-slot of $\prota(n,2^l)$
for some
   $1\le l\le \lceil\log n \rceil$, then it will stay inactive for
all the rest of   $\protb(n)$.

\begin{thm}\label{upper bound-B}
    Protocol  $\protb(n)$  completes broadcasting {\em and terminates}
      in   $O(D \Delta\log n \log (n/\Delta))$ time-slots on any $n$-node
graph   of  maximum eccentricity $D$ and  maximum
   in-degree   $\Delta$.
\end{thm}

\proof
Since $G$ has maximum in-degree $d$, the execution of
$\prota(n,2^{l_d})$ where
$l_d$ is the minimum integer such that $d\le 2^{l_d}$ satisfies Claim  in
the proof of Theorem~\ref{single::upper bound}. Observe that  during  this
execution, a node $v$, that   satisfies the two conditions for
transmitting, could be already in the non active state
because of the execution of some    $\prota(n,2^{l})$ with $l \le
l_d$. However,
if this is the case, $v$ has already successfully transmitted the source
message to
all  its out-neighbors.

\eproof

\noindent
In the third  protocol, the nodes  only  know   their respective
labels.

\smallskip

\noindent
{\bf Description of Protocol \prot{\alpha} ($\alpha > 1$).}
Informally speaking,
this protocol consists in running  $\protb(n)$ with
$n=2^{\ell}$,   for $\ell=1,2,\ldots$
     One of these executions will be the ``good'' one.  However,
     applying a direct ``dovetail'' scheduling would result into
     a completion time of $O\left((D \Delta\log n\log(n/\Delta))^2\right)$
      (recall that nodes do not know $n$).
      So, in order to bound the
     extra-time by a factor of $O(\log^{\alpha} n)$,  Protocol $\prot{\alpha}$
executes   different
     applications of $\protb(\cdot)$ according to a more sophisticated   {\em
     dovetail}
technique.
      Consider the following family of functions:

\[  
f^{\alpha}_0(z)=0 , \:
f^{\alpha}_k(z) \ = \ 2^{\left\lceil k^{\frac{2}{\alpha}}\right\rceil}(k-z), \:
k=1,2,3,\ldots
\]

\begin{figure}[bp]
	\centering
	 \epsfig{scale=0.7,file=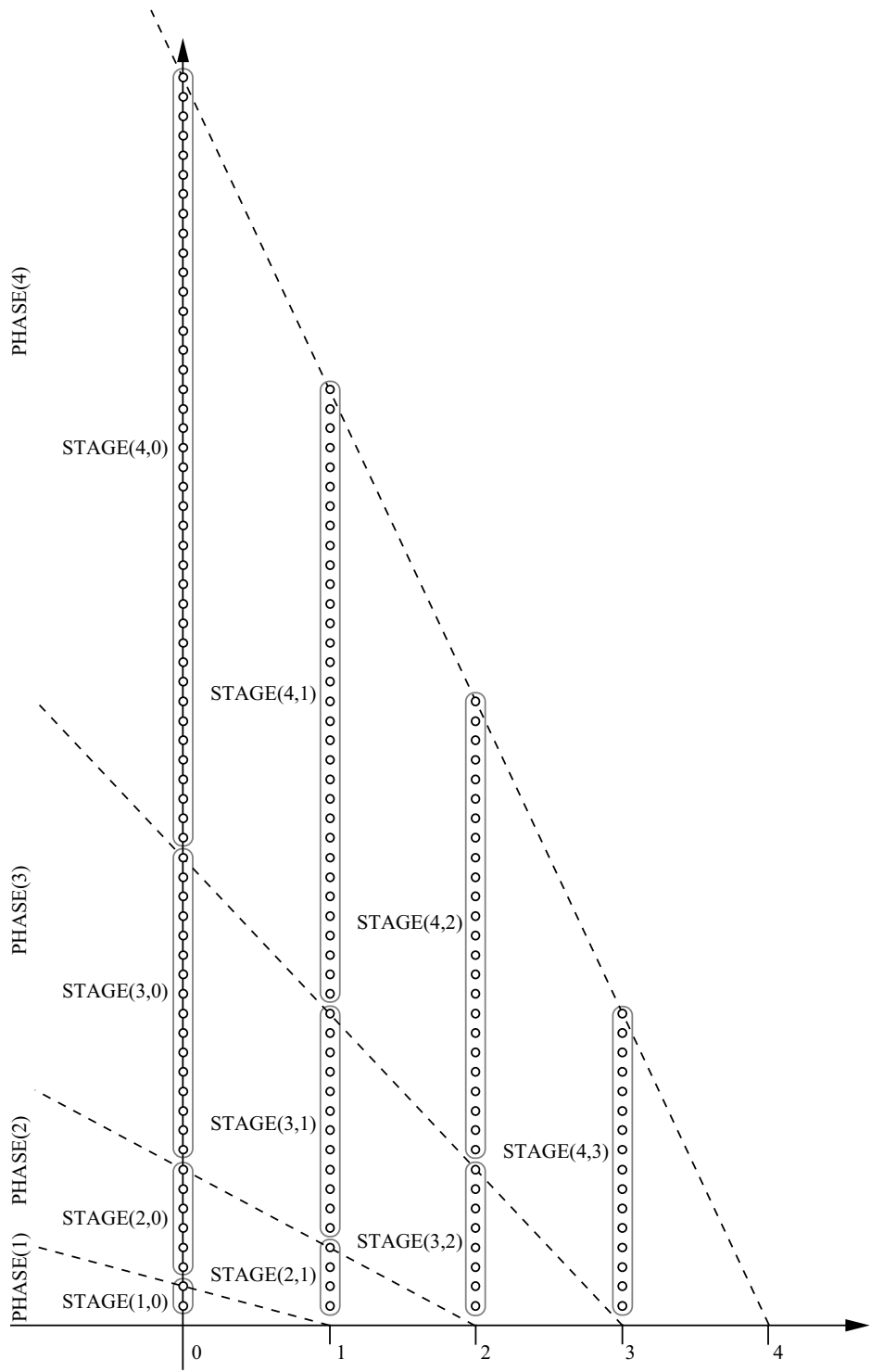}
	 \caption{The figure refers to the case $\alpha = 2$: the abscissa
	represents the executions of $\protb(2^{\ell})$, while the ordinate
	represents the time-slots of $\protb(2^{\ell})$.}
	\label{Fig::dovetail}
\end{figure}

\noindent
Protocol $\prot{\alpha}$ consists of a sequence of {\em phases}, denoted as
$\phase (k)$, $k=1,2,3, \ldots$
The $\phase(k)$ is in turn formed by $k$ {\em
stages}:
in $\stage(k,\ell)$ (with $\ell=0,1, \ldots,k-1$), the  nodes execute the
   time-slots

\[ f^{\alpha}_{k-1}(\ell) + 1, f^{\alpha}_{k-1}(\ell) + 2, \ldots ,
f^{\alpha}_{k}(\ell) \]

\noindent
of    $\protb(2^{\ell})$.
If a node $v$ is not  active   in   a time-slot
     of $\protb(2^{\ell})$,   for some $\ell$, then it will remain  non
      active for all the rest
of   $\prot{\alpha}$. This new dovetail technique is shown in Figure
\ref{Fig::dovetail}. Observe that a node $v$ during the execution
of a time-slot
of $\protb(2^{ \ell})$ could have been informed for the first time
   during a time-slot
of the execution of
    $\protb(2^{ \ell `})$ for some $\ell ` \ne \ell '$. In this case,
     by definition of Protocol
$\prota(\cdot,\cdot)$, the node $v$ acts as an informed node.

\begin{thm}\label{upper bound-C}
    For any positive constant $\alpha >0$,
       $\prot{\alpha}$ completes broadcasting and {\em terminates}
      in   $O(D \Delta\log(n/\Delta) \log^{1+\alpha} n)$ time-slots on 
any $n$-node
graph  of maximum eccentricity $D$ and  maximum in-degree     $\Delta$.
\end{thm}
\proof
   The execution of $\protb(2^{\ell_n})$, for $\ell_n = \lceil
   \log n \rceil$, will be the
   good one and it has completion time $O(D\Delta\log (n/\Delta) \log n )$.
      By definition of     $\prot{\alpha}$,  it   follows that all the nodes
      turn to the non active state within the last time-slot
       $t_{end}=O(D\Delta\log (n/\Delta) \log n)$
of  $\protb(2^{\ell_n})$.  We thus need to upper bound the time in which
      this happens, i.e.,  when $\phase(k_{end})$ is completed,
       where $k_{end}$ is the
smallest integer such that $k_{end} > \ell_n$ and
$ f^{\alpha}_{k_{end}}(\ell_n) \ge t_{end}$.
  From the definition of $f^{\alpha}_{k}$, it follows that

\begin{equation}\label{eq::kf}
   k_{end}  \le \log^{\frac{\alpha}{2}}t_{end}
   \end{equation}

   \noindent
Let $T$ be the number of time-slots required to complete $\phase(k_{end})$.
  From the definition of phase and stage of the protocol, it holds
that

\[ T = \sum_{k=1}^{ k_{end}}\sum_{\ell = 0}^{ k-1}\left(
f^{\alpha}_k(\ell)
   - f^{\alpha}_{k-1}(\ell)\right) =\sum_{\ell = 0}^{k_{end}-1}
   f^{\alpha}_{k_{end}}(\ell)
   \]

\noindent
It thus follows that

\begin{equation}\label{eq::T}
   T = \sum_{\ell = 0}^{k_{end}-1} f^{\alpha}_{k_{end}}(\ell)
    \le 2^{\left\lceil k_{end}^{\frac{2}{\alpha}}\right\rceil}
     \sum_{\ell = 0}^{k_{end}-1} (k_{end}-\ell) \le
2^{\left\lceil k_{end}^{\frac{2}{\alpha}}\right\rceil}
k_{end}^2
   \end{equation}

\noindent
Finally, by combining Ineq.~\ref{eq::kf} with
Ineq.~\ref{eq::T}, we get
$    T \le  t_{end} \log^{\alpha}t_{end}$ and so
   $$T = O(D \Delta\log(n/\Delta) \log^{1+\alpha} n) . $$
\eproof

\section{Multi-Broadcast: The BB model}\label{sec::multy}


  We first need to extend Definition \ref{def::the single model} to
  multi-broadcast
  operations.
  We assume here that each message has an header containing a unique 
  ID number so that two messages have different ID numbers.

   \begin {defi}\label{def::prot}
    A  {\em multi-DB} ({\em multi-RB}) protocol
    $P$ is a   protocol that, given a graph $G$ and a  set of $r$
    broadcast
    operations on it (in short, an $r$-broadcast operation, $r\ge 1$), works
in time-slots  (numbered $0,1,\ldots $) according to the following rules.

    \begin{enumerate}
           \item  In every  time-slot, each node either acts as   transmitter
          or as   receiver.  When transmitting, the node sends   {\em one}
message.

          \item   All the nodes share the same {\em message-recovering}
function
$\sR$ that   takes any message $m$ as input and, if any, returns the
(unique) broadcast message
contained in $m$.

          \item  A node receives a message in a time-slot if and only if
          it acts as  receiver   and {\em exactly} one of its
          $in$-neighbors acts as  transmitter in that time-slot.

          \item  The actions of a node in a specific time-slot are
            function of   its own label, the
          number of the current time-slot $t$, and the messages received
          during  the previous time-slots (for multi-RB protocols, the
          actions also depend on the output of a random bit generators).

\end{enumerate}
\end {defi}

\subsection{The lower bound for deterministic protocols}

\noindent
The next theorem provides  a lower bound which is  a function of the congestion
$c$ and $n$.

    \begin{thm}  \label{thm::lb-1}
        Let $P$ be any multi-DB protocol.
Then, for any    $n\ge 4$ and $c\ge 2$,
      there exist   an $n$-node directed graph $G^P$,    with $D=3$,
    $\Delta=2$, and   an $r$-broadcast operation (with $r\ge c$)
    on $G^P$ (yielding a congestion $c$), such that $P$ completes this operation
    in  $\Omega((c/\log c) \log n)$  time.
    \end{thm}

\proof
Without loss of generality  we consider the case in which $c$ is arbitrary
fixed and $r=c$ (i.e.,
maximum congestion).
   The graph $G^P$   will be one of   the family  $\sG_n$ described below.
   Any     graph in  $\sG_n$ is an $n$-nodes  directed graph of 3 levels. The
first level
consists of  the (unique) source node (with label 1)
   containing all the $r$ messages. The source
   is then connected to $n-2$ nodes that form the second level. Finally,
       the third level has only one node (with label $n$), i.e., the {\em
sink}. The sink has
exactly two in-neighbors among the nodes in the    second level.
We denote by $G_{u,v}$ the graph
        in which the sink has $u$ and $v$ as its in-neighbors. So,

        \[
        \sG_n = \{ G_{u,v} \ | \ u,v \in \{2,3,\ldots, n-1\}  \} \]
        Since the sink cannot send any
information to
       any other node, the execution of $P$,  with respect to any non sink node,
       is the same   for every   graph of the family.   Let $T$
be defined as follows

\[ T = \max_{G_{u,v}\in  \sG_n }\{ t \ | \ P \mbox{ has completion time $t$
on } G_{u,v} \} \]

\noindent
We represent the execution of the first $T$ time-slots of $P$,with
respect to to a
node $v$ in the second level,
   as a sequence $\vec x_v$ over the alphabet $\{ 0,1,\ldots , r\}$  with the
following meaning: $\vec x_{v}(t) = z$ ($z\ge 1$) if, at time-slot $t$,
$v$ sends a message $m$ such  that
$\sR(m)$ is the $z$-th source message, where $\sR$ is the
message-recovering function.
Furthermore,
$ \vec x_v(t) = 0$ if,  at time-slot $t$,
either $v$ acts as receiver or it sends a message $m$ such that
  $\sR(m)$ is not defined. We thus obtain a set $\sD^P$ of
$n-2$ sequences of length $T$ over the alphabet $\{ 0,1,\ldots , r\}$.

   \noindent
We claim that a necessary condition to   complete the $r$-broadcast on every
graph  in $\sG$  is that  {\em any}  two sequences $\vec x_u$ and  $\vec
x_v$,
with   $v\neq u$, must be  {\em $r$-different }, i.e.,
for any element $z\in [r]$ there is a coordinate $t$ for which the set
$\{ \vec x_u(t),x_v(t) \}$ is equal to $\{z,0\}$.

\noindent
Indeed, assume by contradiction that this is not true. So, there are two
    sequences
    $\vec x_u$, $ \vec x_v$ with    $v\neq u$   and $z\in [r]$, such that
     $\{ \vec x_u(t),x_v(t) \}$ is not equal to $\{z,0\}$ for every $t\le T$.
     We then  consider
protocol $P$ on the graph $G_{u,v}$. It is easy to verify that, in
$G_{u,v}$,
   the sink is reachable from the source, but the sink  does  not
   receive the $z$-th message during the first $T$ time-slots.

\noindent
  From the above discussion, the $n-2$ sequences of length $T$
   in $\sD^P$ must be pairwise
$r$-different.
  From Theorem~\ref{intro:lem:r-diff}, in order to
have  a set of $n-2$  pairwise $r$-difference sequences of length $T$, it
must hold that
$$ T = \Omega \left (\frac{r}{\log r}\log n \right) $$

\noindent
Since $T$ is a lower bound on the worst-case completion time of $P$
over the graph family $\sG_n$, the theorem follows.
\eproof

\begin{thm} \label{intro::thm::lb-1}
Let $P$ be any multi-DB protocol. Then, for any
$n\ge 4$, $3\le D \le n/6$  and $2 \le c\le r$,
it is possible to define
      an $n$-node directed graph $G^P$ with maximum
    eccentricity $D$ and  maximum in-degree
    $\Delta=2$, and  a set of $r$ independent broadcast   operations
    on $G^P$ (yielding a congestion $c$) such that $P$ completes these
operations on $G^P$ in
     $\Omega(c + (c/\log c + D) \log n)$ time.
    \end{thm}
\proof
Since a node can receive at most
one message per time-slot, it is easy to obtain the lower bound $\Omega(c)$.
Then, by combining the family of graphs of
  Theorem ~\ref{cor::LOW.BOUND-d} (with $\Delta=2$) with
  that yielding Theorem~\ref{thm::lb-1}, we
easily get the thesis.
\eproof

\noindent
Notice that the proof of   Theorem~\ref{thm::lb-1} does not rely on the fact
that
the nodes do not know $n$. Thus,   Theorem \ref{intro::thm::lb-1} 
also holds under
this condition.

\noindent
Finally, we observe that the above contruction can be easily modified in
order to
let  each of the $r$ broadcast messages start from a different source
(i.e.
we have $r$ messages in $r$ different sources). It suffices to replace
the source node in $\sG_n$ (see the proof of Theorem \ref{thm::lb-1})
with the root of
a binary directed  tree in which the
$r$ sources are the leaves of the tree. The only difference is
that $D$ is now a logarithmic function of $n$.

\subsection{Lower bound for randomized protocols}

Any worst-case time bound of  a randomized protocol can be considered
reliable if it  happens within a high probability on every possible instance.
This  concept is widely adopted in the field of randomized
algorithms~\cite{MR95}, and
it can be easily adapted to the case of multi-broadcast operations on
unknown networks.

\begin{defi}\label{def::rand-prot}
A multi-RB protocol $P$ has {\em reliable} completion time
$T$ (where, clearly, $T$ depends on $n$)  if, for any $n\ge 1$
   and  for any $n$-node graph $G$, $P$
completes, with probability at least  $1-1/n$
   any $r$-broadcast operation on $G$ within $T$ time-slots.
\end{defi}

\begin{lem}\label{rlb}
  Let $P$ be a multi-RB protocol for unknown networks.
If $P$ has reliable completion time $T$, then it must holds that
$T  = \Omega((c/\log c) \log n)$.
\end{lem}

\proof
The proof makes use of the
families $\sG_n$ ($n\ge 4$) of directed graphs, and the corresponding
multi-broadcast operations,
which have been introduced in the proof of Theorem~\ref{thm::lb-1}.
   In particular, we
will show  that, for any $n\ge 4$, a graph $G^P \in \sG_n$ exists on which
$P$ has   $\Omega((c/\log c) \log n)$ completion time  with probability
larger than  $1/n$. As in the proof of Theorem~\ref{thm::lb-1}, an execution
of $T$ time-slots of $P$, with respect to  the nodes of the second level of
$G_{u,v}$, can be represented as a set $\sD$ of $n-2$ sequences of length
$T$ over the alphabet $\{0, \ldots, r\}$ (with the same meaning of that
given in the proof of Theorem~\ref{thm::lb-1}).
A multi-RB protocol (restricted to
   the   nodes in the second level of $G_{u,v}$)
can   thus be seen as a probability distribution $\sP$ over the set
$\sA$ of all possible sequence sets $\sD$.
Consider the following function

\[
\chi^{\sD}(u,v) = \left\{
\begin{array}{ll}
	1 \ \  & \mbox{ if  sequences } u   \mbox{ and } v \mbox{ in } {\sD}
	\mbox{ are } r\mbox{-different} \\
	0 \ \ & \mbox{ otherwise }
\end{array} \right.
\]

\noindent
 From the hypothesis of the theorem we have that, for all
$u\neq v$,

\[
\Pr\{ P \mbox{ has completion time $T$ on }
G_{u,v} \} \ge 1 - \frac 1n, \]
it follows that
\begin{equation}\label{Eq::prob-1}
	\forall u\neq v\;\; \sum_{{\sD} \in \sA} 
\sP({\sD})\chi^{\sD}(u,v) \ge 1 - \frac 1n
\end{equation}

   \noindent
   Consider now the sum

   \[ M^{\sD} = \sum_{u,v \in L_2} \chi^{\sD}(u,v) \]
\noindent where $L_2$ denotes the nodes of the second  level of any
graph in  $\sG_n$;
this equals the number of $r$-different pairs yielded by the
   protocol $P$. We now prove that
there has to exist a  $\overline{\sD}$ such that

    \begin{equation}\label{Eq::prob-2}
     M^{\overline{\sD}} \ge \left(1-\frac 1n \right)\frac{(n-2)(n-3)}{2}
      \end{equation}

     \noindent
    In fact, from Ineq. \ref{Eq::prob-1}, we have that

    \[ \sum_{u,v \in L_{2}} \sum_{\sD \in \sA} \sP(\sD) 
\chi^{\sD}(u,v) \ge \left(1-\frac
    1n\right)\frac{(n-2)(n-3)}{2} \]
and, hence
\[
\sum_{\sD \in  \sA} \sP(\sD) \sum_{u,v\in L_{2}}\chi^{\sD}(u,v) \ = \
\sum_{\sD \in \sA} \sP({\sD}) M^\sD
   \ \ge \ \left(1-\frac 1n\right)\frac{(n-2)(n-3)}{2}
\]
It follows that a   $\overline {\sD} $ exists   that verifies
    Ineq.ó\ref{Eq::prob-2}.
   We now prove that $\overline {\sD}$ must contain a large subset of
    sequences which are pairwise $r$-different, thus the same
    property derived in the proof of Theorem~\ref{thm::lb-1}.

\medskip

\begin{verse}
{\bf Claim.} There exists a subset  $\overline\sS \subseteq \overline {\sD}$
such that  $|\overline\sS | = n/2$ and, for any $u,v \in \overline\sS $ (with
$u \neq v$), the pair $(u,v)$ is $r$-different.

\noindent {\bf Proof.}
    The following simple
    algorithm   finds the desired subset $\overline\sS $ (we assume here that
    $n$ is an even number).

\begin{small}
\begin{itemize}
   \item[] {\bf begin}
   \item[] Choose     an arbitrary $\overline\sS \subseteq \overline {\sD}$ s.t.
   $|\overline\sS | = n/2$
\item[]    $\overline\sS^c := \emptyset$:

         \item[] {\bf while} ($\overline\sS$ does not satisfy the 
claim) {\bf do}

\begin{itemize}
   \item[] {\bf begin} \\
          Choose (arbitrarily) $u,v \in \overline\sS$ that are not 
$r$-different\\
          Choose (arbitrarily) $k \in  \overline{\sD} \setminus 
(\overline\sS \cup
          \overline\sS^c)$\\
           $\overline\sS := (\overline\sS - \{ v \}) \cup \{k \}$;\\
           $\overline\sS^c := \overline\sS^c \cup \{ v \} $\
         \item[] {\bf end}
         \end{itemize}

         \item[]  {\bf return} $\overline\sS$.
         \item[] {\bf end}

\end{itemize}

\end{small}

\noindent
   We first notice that, from Ineq.ó\ref{Eq::prob-2},
    there are at most $(n-3)/2$ different
    pairs in $\overline {\sD}$ that are not $r$-different. Since at every
    iteration of the while loop the algorithm ''discards''
    a new not $r$-different pair in $\overline {\sD}$, the algorithm
     always returns a set $\overline\sS$ satisfying the claim in
     $O(n)$  steps.
\end{verse}

\noindent The claim  implies that the reliable multi-RB protocol
    $P$ on the family $\sG_n$ must yield a set of $n/2$ sequences that
    are pairwise $r$-different. So, by applying the lower bound
    of Theorem~\ref{intro:lem:r-diff}, we can state that

    \[ T = \Omega\left( \frac r{\log r}\log \frac n2 \right)
    = \Omega\left( \frac r{\log r}\log n \right) \]
    Since $T$ is a lower bound on the completion time
    of $P$, the lemma follows.
  \eproof

   The proof of the following theorem is an easy consequence
of the $\Omega(D\log (n/D))$ lower bound for randomized protocols
given in \cite{KM93}, the trivial lower bound $\Omega(c)$,
and Lemma~\ref{rlb}.

\begin{thm}\label{intro::rand-low-bound}
   Let $P$ be any multi-RB protocol. Then, for any
        $n\ge 4$, $D\ge 3$  and $2\le c\le  r$,
    there exist (i)  an $n$-node directed graph $G^P$ with maximum
    eccentricity $D$ and  maximum in-degree
    $\Delta=2$, and (ii) a  set of $r$ independent
    broadcast operations
    on $G^P$ (yielding a congestion $c$), such that the   reliable completion
time
    of  $P$        is
     $\Omega(c+ (c/\log c)\log n +D \log (n/D))$.
\end{thm}

The above theorem
   in fact holds for  any  probability lower bound of
the form $1-1/n^a$ (with any fixed constant $a>0$). However, we
choose the form $1-1/n$ in order to simplify   the   proof
of Lemma \ref{rlb}.

    \subsection{The multi-broadcast protocols}\label{sec::UPPER BOUNDS}

   As mentioned in the Introduction,
   our multi-DB protocols make use of superimposed
   codes \cite{EFF,HS,I97,CHI99}.
%
%
In particular, we will use the following upper bounds \cite{EFF,KS64}.

\begin{thm}\label{sec:th::upperbound-ssel}
For any  $n\ge 3$ and for $k \ge 2$,
\begin{description}
\item[]\cite{EFF}
there exist ($n,k$)-strongly-selective families of size
   $O(\min\{n,k^2\log n\})$;
\item[]\cite{KS64}(p. 370) it is possible to construct, in  time polynomial
in $n$ and $k$, an ($n,k$)-strongly-selective family (based on $q$-ary
{\em error-correcting codes})
   of size $O(\min\{n,k^2\log^{2} n\} )$.
\end{description}
\end{thm}
\noindent We recall that  an $\Omega((k^{2}/\log k)\log n)$
lower bound is proved in Theorem \ref{th:strongly-low}.

In what follows, we describe the protocol
$\allprota(n,\Delta)$: It  assumes that nodes    know   
$\Delta$ and     $n$.  However,
when $\Delta$ and/or $n$ are not known, we can adopt the same
dovetail technique described in subsection \ref{upp_b}. The cost of 
this  further
task
is $O(\log^{1+\alpha}n)$ additional time-slots (for any fixed $\alpha >0$)
per each time-slot of $\allprota(n,\Delta)$.

\medskip

\noindent
{\bf Description of Protocol $\allprota(n,\Delta)$.}
With each message is associated a priority so that the priorities induce
a total ordering on the set of broadcast messages\footnote{A possible
choice for the priorities is the following. If $m$ is the $h$-th message
of a source of label $l$ then the priority of $m$ is given by the pair
$(l,h)$. Thus, for any two messages $m$ and $m'$ of priorities $(l,h)$ and
$(l',h')$, $m$ has priority higher than $m'$ if either $l<l'$ or $l=l'$
and $h<h'$.}. The priorities will be used by the nodes to schedule  the
messages to send. In fact, every node stores the messages by means of a
priority
queue. The protocol $\allprota(n,\Delta)$ uses an
$(n,\Delta+1)$-strongly-selective family
$\sF=\{F_1,F_2, \ldots F_{|\sF|}\}$. It consists of a sequence of
consecutive {\em phases}. Each phase consists of $|\sF|$ time-slots.
   At the very beginning, the priority queues of the source nodes contain
their
   broadcast messages, the other priority queues are empty. At the
   beginning of each phase, every node $v$ with a non empty priority
   extracts from the queue the message $m_v$  of highest priority. At the
   $j$-th time-slot of the phase,  node $v$ acts according to the following
   rules.

\begin{verse}
   - If the label of $v$ belongs to $F_j$ and $m_v$ exists then $v$ transmits
   $m_v$. \\
   - In all the other cases, $v$ acts as a receiver. If $v$ receives a
   message that $v$ has never received before, the message is enqueued,
   otherwise, the message is discarded.
   \end{verse}

   \begin{thm}\label{upper bound}
Protocol  $\allprota(n,\Delta)$  completes any $r$-broadcast operation
within  $O((D+c) \min\{n,\Delta^2\log n\})$ time-slots on  any $n$-node
graph of maximum eccentricity $D$,  maximum in-degree     $\Delta$, and
congestion $c$.
\end{thm}

\proof
Firstly, we prove the following

\medskip

\begin{verse}
{\bf Claim 1}\quad
If a node $v$ transmits a message during a phase $t$ then
all the out-neighbours of $v$ receive the message by the
end of phase $t$.

\noindent {\bf Proof.} In fact, let $u$ be any out-neighbour of $v$.
By virtue of the $(n, \Delta+1)$-strongly selectivity of $\sF$,
there is a time-slot of phase $t$ in which $u$ acts as a receiver and
$v$ is the only node, among the in-neighbors of $u$ (notice that
these are at most $\Delta$),
that  transmits. Hence, in that time-slot, $u$ receives
the message of $v$.

\end{verse}

\medskip

\noindent Now, we can show that all   messages reach their destinations.

\medskip

\begin{verse}

{\bf Claim 2}\quad
After a finite number of phases, any message $m$ is received by
all the nodes that are reachable from the source of $m$.

\noindent {\bf Proof.} The proof is by induction on the distance $\ell$ from
the source.
For $\ell = 0$ the claim is obvious. Let $u$ be a node at distance
$\ell + 1$ from the source. Consider an in-neighbour $v$ of $u$ at
distance $\ell$. From the inductive hypothesis, node $v$ receives the
message $m$. The first time this happens, $m$ is enqueued in
the priority queue of $v$. According to the protocol,
at the beginning of each phase, node $v$ extracts the message
of highest priority and transmits it. Since there are less than
$r$ messages of priority higher than that of $m$ and any message is
enqueued at most once, it follows that $v$ extracts and transmits
$m$ within at most $r$ phases. Thus, from Claim~1 node $u$
receives the message $m$.

\end{verse}

\medskip

\noindent The last step consists of showing that the messages cannot
be delayed too much.

\medskip

\begin{verse}
{\bf Claim 3}
Assume that node $u$, at distance $\ell$
from the source of a message $m$, receives $m$
   for the first time during phase $\ell + t_u$.
Then, during the first $\ell+t_u$ phases, $u$ transmits at
least $t_u -1$ messages with higher priority than that of $m$.

\noindent {\bf Proof.} The proof is by induction on the
distance $\ell$. For $\ell=0$ the claim is obvious. Let $u$ be a node at
distance $\ell + 1$ and let $v$ be the node
at distance $\ell$ from which $u$ receives $m$ for the
first time. Let $\ell +t_v$ be the phase in which node
$v$ receives $m$ for the first time.
By inductive hypothesis,
in the first $\ell + t_v$ phases, $v$ transmits at least $t_v -1$ messages
of higher priority (than that of $m$). This fact and Claim~1 imply
that, in the first $\ell+t_v$ phases, $u$
receives from $v$ a set $M_1$ of at least $t_v -1$ messages of higher
priority.
Now, let $k\geq 0$ be the number of phases during which the
message $m$ remains in the priority queue of $v$.
This implies that $v$ transmits a set $M_2$ of $k$ further messages
of higher priority before transmitting
$m$. As a consequence, $u$ receives all the messages in $M_2$
before receiving $m$.\\
\noindent Let $t_u = t_v  + k$; then
node $u$ receives $m$ for the first time in
phase $\ell + 1 + t_u$.
When this happens $u$ has received from $v$ at least
$t_v - 1 + k = t_u - 1$ messages of higher priority (i.e., all the messages
in $M_1\cup M_2$).
Since all these messages have been received in distinct phases among the
first $\ell+ t_u$, the node $u$ transmits at least $t_u -1$ messages of
higher priority in the first $\ell + 1 + t_u$ phases.

 \end{verse}

\medskip

\noindent Notice that any message $m$ has to reach nodes at distance at most
$D$ from its source and there are    less than $c$ messages of
priority higher than $m$ that collide with $m$. Hence, Claim~2 and
Claim~3 imply that any message reaches its
destinations in less than $D+c$ phases. Since each phase requires
$O( \min\{n,\Delta^2\log n\})$ time-slots (see Theorem.~\ref{sec:th::upperbound-ssel})
the thesis follows.
\eproof

   \section{Conclusions and open problems} \label{sec::CONCLUSIONS}

The main contribution of this paper is that of
providing   explicit and strong connections between a set of old and
new
combinatorial tools  and the issue of broadcast
operations in
radio  networks of unknown topology. Thanks to such connections,
we have  obtained new lower and upper bounds on the completion time
of  this important operation. We believe that the concept of
selectivity and that of $r$-different
sequences can have further applications to other
distributed models in which the local knowledge is extremely low.
Some evidence of our opinion is also  given by some
previous results  \cite{Li92} in which strongly-selective
families have been used for the distributed coloring problem
in unknown graphs.

As for specific research directions, the following ones  appear to be
the most relevant.

\noindent
The upper bounds for the DB protocols are not
efficiently constructible since the corresponding ``small'' selective families
have been derived by using probabilistic arguments: it would thus be
  important to design an efficient deterministic algorithm that
constructs such families.

  \noindent
    The lower bounds for multi-DB   (and multi-RB)  protocols  are
   consequences of the combinatorial
    lower bound $\Omega(\frac{r}{\log r}\log n)$  on the length of $n$
     pairwise $r$-different sequences given in   Theorem~\ref{intro:lem:r-diff}.
      We do not know
    whether the latter  is tight. As far as we know, the
    best upper bound is $O(r\log n)$.
The deterministic upper bounds in Theorem~\ref{upper bound}
    almost match
   the lower bound in Theorem~\ref{intro::thm::lb-1}, when $\Delta=O(\polylog 
n)$.
    An interesting
   future research goal is that of reducing the gap between upper and
   lower bounds for larger   $\Delta$. To this aim, we believe
   that a  generalization of   the $\Omega(\frac{r}{\log r}\log n)$ lower bound
   to $d$-wise $r$-different sequences  could give
   a stronger lower bound that also depends   on $\Delta$.

A further research direction is that of using the same combinatorial
tools to investigate the issue of {\em dynamical-fault tolerance} in radio networks.
Dynamical edge and node faults     may   happen   at
  any instant, {\em even during the execution of a protocol}. 
Some results on this direction have been recently  obtained in
\cite{CMS01c}.

   \medskip
   \noindent
   {\bf Acknowledgments.} We wish to thank
   J\`anos K\"orner  for helpful discussions.

\end{document}